\documentclass[12pt]{article}
\usepackage{amssymb,amsfonts}
\usepackage{epsf,epsfig}
\begin{document}
\baselineskip 18pt

\title{Bethe Ansatz approach to the three-magnon problem in an infinite $XX$ spin chain with Izing next neighboring ferromagnetic interaction}
\author{P.~N.~Bibikov}
\date{\it Russian State Hydrometeorological University, Saint-Petersburg, Russia}
\maketitle

\vskip5mm

\begin{abstract}
Using the traditional and degenerative discrete-diffractive versions of Bethe Ansatz we construct a rude set of three-magnon states in the infinite $XX$ spin chain with Izing next neighboring ferromagnetic term.
\end{abstract}

\maketitle

\section{Introduction}

For a long time it was a common opinion that the Bethe ansatz works only for integrable models \cite{1,2,3,4}. However
as it was recently shown \cite{5} the modified version of the Bethe ansatz, which will be called the degenerative
discrete-diffractive (DDD) Bethe ansatz may be applied for solution of the {\it three-magnon} problem in a {\it non-integrable} spin model defined on the {\it infinite} 1D lattice if the corresponding two-magnon scattering matrix satisfies the factorization condition \cite{6,7}. An example of such class of models is the $S=1$ isotropic ferromagnetic infinite chain \cite{5}. In the present paper we apply the suggested approach to the Hamiltonian
\begin{equation}
\hat H=-\sum_{n=-\infty}^{\infty}\Big[\frac{J}{2}\Big({\bf S}^+_n{\bf S}^-_{n+1}+{\bf S}^-_n{\bf S}^+_{n+1}\Big)+
J_z\Big({\bf S}^z_n{\bf S}^z_{n+2}-\frac{1}{4}\Big)+{\gamma h}\Big({\bf S}_n^z-\frac{1}{2}\Big)\Big],\quad J\neq0.
\end{equation}
Here ${\bf S}_n$ is the standard triple of spin-1/2 operators related to the $n$-th site. For appropriate values of the coupling parameters (for example for rather big $\gamma h\geq0$) this system has the ferromagnetic zero-energy ground state
\begin{equation}
|\emptyset\rangle=\prod_{n=-\infty}^{\infty}\otimes|\uparrow\rangle_n,
\end{equation}
where $|\uparrow\rangle_n$ and $|\downarrow\rangle_n$ is the standard doublet of spin-1/2 states. Hamiltonian (1) is the simplest nontrivial example in the large list of models with nearest neighbor and over nearest neighbor interactions for which the two-magnon spectrum may be evaluated \cite{8}.

The corresponding one-magnon states are the flat waves
\begin{equation}
|magn,k\rangle=\sum_n{\rm e}^{ikn}{\bf S}^-_n|\emptyset\rangle,
\end{equation}
related to the energy
\begin{equation}
E_{magn}(k)=J_z+\gamma h-J\cos{k}.
\end{equation}

The two-magnon problem may be well studied in the standard manner. The corresponding results will be presented in the next section. The three-magnon problem is more complicated because the system (1) is non-integrable. It will be considered in Sect. 3.

\section{Two-magnon states}

For a two-magnon state
\begin{equation}
|2\rangle=\sum_{n_1<n_2}a_{n_1,n_2}{\bf S}^-_{n_1}{\bf S}^-_{n_2}|\emptyset\rangle,
\end{equation}
the wave function satisfies the ${\rm Schr\ddot odinger}$ equation in the form of the following system
\begin{eqnarray}
&&2(J_z+\gamma h)a_{n_1,n_2}-\frac{J}{2}\sum_{l=\pm1}\Big(a_{n_1+l,n_2}+a_{n_1,n_2+l}\Big)=Ea_{n_1,n_2},\quad
n_2-n_1>2,\\
&&(J_z+2\gamma h)a_{n,n+2}-\frac{J}{2}\Big(a_{n-1,n+2}+a_{n+1,n+2}+a_{n,n+1}+a_{n,n+3}\Big)=Ea_{n,n+2},\nonumber\\
&&2(J_z+\gamma h)a_{n,n+1}-\frac{J}{2}\Big(a_{n-1,n+1}+a_{n,n+2}\Big)=Ea_{n,n+1}.
\end{eqnarray}
Its solution may be obtained by various approaches \cite{5,9} and has the form
\begin{eqnarray}
&&a_{n_1,n_2}(k_1,k_2)=\sum_{a,b=1}^2\varepsilon_{ab}\Big[A(k_a,k_b)(1-\delta_{n_2-n_1,1})+
B(k_a,k_b)\delta_{n_2-n_1,1}\Big]{\rm e}^{i(k_an_1+k_bn_2)},\nonumber\\
&&E(k_1,k_2)=E_{magn}(k_1)+E_{magn}(k_2),
\end{eqnarray}
where $\varepsilon_{ab}$ is the Levi-Civita tensor and
\begin{equation}
A(k,\tilde k)=J\cos{\frac{k+\tilde k}{2}}-2J_z{\rm e}^{i(k-\tilde k)}\cos{\frac{k-\tilde k}{2}},\qquad
B(k,\tilde k)=J\cos{\frac{k+\tilde k}{2}}.
\end{equation}
Really substitution (8) solves automatically Eq. (6) and reduces Eqs. (7) to the system
\begin{eqnarray}
&&\sum_{a,b=1}^2\varepsilon_{ab}\Big[\frac{J}{2}{\rm e}^{ik_b}\Big(1+{\rm e}^{i(k_a+k_b)}\Big)
\Big(A(k_a,k_b)-B(k_a,k_b)\Big)-J_z{\rm e}^{2ik_b}A(k_a,k_b)\Big]=0,\\
&&\sum_{a,b=1}^2\varepsilon_{ab}\Big[B(k_a,k_b)(\cos{k_a}+\cos{k_b}){\rm e}^{ik_b}-\frac{1}{2}A(k_a,k_b)
\Big({\rm e}^{i(k_b-k_a)}+{\rm e}^{2ik_b}\Big)\Big]=0,
\end{eqnarray}
which may be readily solved by the substitution (9).

As in the case of the $S=1$ ferromagnetic chain \cite{5} the two-magnon S-matrix (which is $S(k_1,k_2)=A(k_2,k_1)/A(k_1,k_2)$) is $1\times1$-dimensional and hence automatically satisfies the factorization condition \cite{6,7} (the Yang-Baxter equation).

\section{The Three-magnon problem}

\subsection{The standard Bethe Ansatz approach}

For the wave function of a three-magnon state
\begin{equation}
|3\rangle=\sum_{n_1<n_2<n_3}a_{n_1,n_2,n_3}{\bf S}^-_{n_1}{\bf S}^-_{n_2}{\bf S}^-_{n_3}|\emptyset\rangle,
\end{equation}
the corresponding ${\rm Schr\ddot odinger}$ equation splits on two groups of equations. The former one
\begin{eqnarray}
&&[(3-\delta_{n_2-n_1,2}-\delta_{n_3-n_2,2})J_z+3\gamma h]a_{n_1,n_2,n_3}-\frac{J}{2}\sum_{l=\pm1}\Big(a_{n_1+l,n_2,n_3}\nonumber\\
&&+a_{n_1,n_2+l,n_3}+a_{n_1,n_2,n_3+l}\Big)=Ea_{n_1,n_2,n_3},\qquad
n_2-n_1>1,\quad n_3-n_2>1,\nonumber\\
&&3(J_z+\gamma h)a_{m,n,n+1}-\frac{J}{2}\Big(a_{m-1,n,n+1}+a_{m+1,n,n+1}\nonumber\\
&&+a_{m,n-1,n+1}+a_{m,n,n+2}\Big)=Ea_{m,n,n+1},\qquad n-m>2,\nonumber\\
&&3(J_z+\gamma h)a_{m-1,m,n}-\frac{J}{2}\Big(a_{m-2,m,n}+a_{m-1,m+1,n}\nonumber\\
&&+a_{m-1,m,n-1}+a_{m-1,m,n+1}\Big)=Ea_{m-1,m,n},\qquad n-m>2,
\end{eqnarray}
is related to free motion and binary collisions while the latter one
\begin{eqnarray}
&&(2J_z+3\gamma h)a_{n-1,n,n+2}-\frac{J}{2}\Big(a_{n-2,n,n+2}+a_{n-1,n+1,n+2}+a_{n-1,n,n+1}\nonumber\\
&&+a_{n-1,n,n+3}\Big)=Ea_{n-1,n,n+2},\nonumber\\
&&(2J_z+3\gamma h)a_{n-2,n,n+1}-\frac{J}{2}\Big(a_{n-3,n,n+1}+a_{n-1,n,n+1}+a_{n-2,n-1,n+1}\nonumber\\
&&+a_{n-2,n,n+2}\Big)=Ea_{n-2,n,n+1},\nonumber\\
&&(2J_z+3\gamma h)a_{n-1,n,n+1}-\frac{J}{2}\Big(a_{n-2,n,n+1}+a_{n-1,n,n+2}\Big)=Ea_{n-1,n,n+1},
\end{eqnarray}
corresponds to three-magnon collisions.

A substitution
\begin{eqnarray}
&&a_{n_1,n_2,n_3}(k_1,k_2,k_3)=\sum_{a,b,c=1}^3\varepsilon_{abc}\Big[A(k_a,k_c)\Big(A(k_a,k_b)A(k_b,k_c)
(1-\delta_{n_2-n_1,1})(1-\delta_{n_3-n_2,1})\nonumber\\
&&+B(k_a,k_b)A(k_b,k_c)\delta_{n_2-n_1,1}(1-\delta_{n_3-n_2,1})
+A(k_a,k_b)B(k_b,k_c)(1-\delta_{n_2-n_1,1})\delta_{n_3-n_2,1}\Big)\nonumber\\
&&+C_{abc}(k_1,k_2,k_3)\delta_{n_2-n_1,1}\delta_{n_3-n_2,1}\Big]{\rm e}^{i(k_an_1+k_bn_2+k_cn_3)},
\end{eqnarray}
with some indefinite functions $C_{abc}(k_1,k_2,k_3)$ solves the system (13) (after reduction it to Eqs. (10), (11)), giving  the usual expression for the energy
\begin{equation}
E(k_1,k_2,k_3)=\sum_{j=1}^3E_{magn}(k_j).
\end{equation}
At the same time Eqs. (14) turn into
\begin{equation}
X^{(j)}(k_1,k_2,k_3){\rm e}^{i(k_1+k_2+k_3)n}=0\Longrightarrow X^{(j)}(k_1,k_2,k_3)=0,\qquad j=1,2,3,
\end{equation}
where
\begin{eqnarray}
&&X^{(1)}(k_1,k_2,k_3)=\sum_{a,b,c=1}^3\varepsilon_{abc}\Big[A(k_a,k_c)\Big(B(k_a,k_b)A(k_b,k_c)
\Big(\frac{J}{2}{\rm e}^{i(k_c-k_a)}\nonumber\\
&&+J(\cos{k_a}+\cos{k_b}){\rm e}^{i(2k_c-k_a)}-J_z{\rm e}^{i(2k_c-k_a)}\Big)-\frac{J}{2}\Big(A(k_a,k_b)A(k_b,k_c){\rm e}^{2i(k_c-k_a)}\nonumber\\
&&+A(k_a,k_b)B(k_b,k_c){\rm e}^{i(2k_c+k_b-k_a)}\Big)\Big)-\frac{J}{2}C_{abc}(k_1,k_2,k_3){\rm e}^{i(k_c-k_a)}\Big],\\
&&X^{(2)}(k_1,k_2,k_3)=\sum_{a,b,c=1}^3\varepsilon_{abc}\Big[A(k_a,k_c)\Big(A(k_a,k_b)B(k_b,k_c)
\Big(\frac{J}{2}{\rm e}^{i(k_c-k_a)}\nonumber\\
&&+J(\cos{k_b}+\cos{k_c}){\rm e}^{i(k_c-2k_a)}-J_z{\rm e}^{i(k_c-2k_a)}\Big)-\frac{J}{2}\Big(A(k_a,k_b)A(k_b,k_c){\rm e}^{2i(k_c-k_a)}\nonumber\\
&&+B(k_a,k_b)A(k_b,k_c){\rm e}^{i(k_c-k_b-2k_a)}\Big)\Big)-\frac{J}{2}C_{abc}(k_1,k_2,k_3){\rm e}^{i(k_c-k_a)}\Big],\\
&&X^{(3)}(k_1,k_2,k_3)=\sum_{a,b,c=1}^3\varepsilon_{abc}{\rm e}^{i(k_c-k_a)}\Big[\frac{J}{2}A(k_a,k_c)
\Big(A(k_a,k_b)B(k_b,k_c){\rm e}^{-ik_a}\nonumber\\
&&+B(k_a,k_b)A(k_b,k_c){\rm e}^{ik_c}\Big)
+\Big(J_z-J(\cos{k_a}+\cos{k_b}+\cos{k_c})\Big)C_{abc}(k_1,k_2,k_3)\Big].
\end{eqnarray}
With the use of Eq. (11) we may rewrite Eqs. (18) and (19) in slightly more compact forms
\begin{eqnarray}
&&X^{(1)}(k_1,k_2,k_3)=\sum_{a,b,c=1}^3\varepsilon_{abc}\Big[A(k_a,k_c)\Big(B(k_a,k_b)A(k_b,k_c)
\Big(\frac{J}{2}{\rm e}^{i(k_c-k_a)}-J_z{\rm e}^{i(2k_c-k_a)}\Big)\nonumber\\
&&+\frac{J}{2}A(k_a,k_b)\Big(A(k_b,k_c)-B(k_b,k_c)\Big){\rm e}^{i(2k_c+k_b-k_a)}\Big)
-\frac{J}{2}C_{abc}(k_1,k_2,k_3){\rm e}^{i(k_c-k_a)}\Big],\nonumber\\
&&X^{(2)}(k_1,k_2,k_3)=\sum_{a,b,c=1}^3\varepsilon_{abc}\Big[A(k_a,k_c)\Big(A(k_a,k_b)B(k_b,k_c)
\Big(\frac{J}{2}{\rm e}^{i(k_c-k_a)}-J_z{\rm e}^{i(k_c-2k_a)}\Big)\nonumber\\
&&+\frac{J}{2}\Big(A(k_a,k_b)-B(k_a,k_b)\Big)A(k_b,k_c){\rm e}^{i(k_c-k_b-2k_a)}\Big)-
\frac{J}{2}C_{abc}(k_1,k_2,k_3){\rm e}^{i(k_c-k_a)}\Big].
\end{eqnarray}
As it will be seen now it is better to use except (20) and (21) their linear combinations
\begin{eqnarray}
&&X^{(+)}(k_1,k_2,k_3)=X^{(1)}(k_1,k_2,k_3)+X^{(2)}(k_1,k_2,k_3),\nonumber\\
&&X^{(0)}(k_1,k_2,k_3)=X^{(3)}(k_1,k_2,k_3)+\frac{J_z-Jc}{J}X^{(+)}(k_1,k_2,k_3),\nonumber\\
&&X^{(-)}(k_1,k_2,k_3)=i(X^{(1)}(k_1,k_2,k_3)-X^{(2)}(k_1,k_2,k_3)).
\end{eqnarray}
A direct calculation with the use of computer algebra system MAPLE gives
\begin{eqnarray}
&&X^{(+)}(k_1,k_2,k_3)=J\Big(\varphi(k_1,k_2,k_3)\Phi^{(+)}(s,c,k)-Z(k_1,k_2,k_3)\Big),\\
&&X^{(0)}(k_1,k_2,k_3)=J_z^2\varphi(k_1,k_2,k_3)\Phi^{(0)}(s,c,k),\nonumber\\
&&X^{(-)}(k_1,k_2,k_3)=-iJJ_z^2\varphi(k_1,k_2,k_3)\Phi^{(-)}(s,c,k).
\end{eqnarray}
Here
\begin{eqnarray}
&&c=\cos{k_1}+\cos{k_2}+\cos{k_3}=\frac{3(J_z+\gamma h)-E(k_1,k_2,k_3)}{J},\nonumber\\
&&s=\sin{k_1}+\sin{k_2}+\sin{k_3},\\
&&\varphi(k_1,k_2,k_3)=\sum_{a,b,c=1}^3\varepsilon_{abc}{\rm e}^{i(k_c-k_a)}=8i\sin{\frac{k_2-k_1}{2}}\sin{\frac{k_3-k_2}{2}}\sin{\frac{k_1-k_3}{2}},\nonumber\\
&&Z(k_1,k_2,k_3)=\sum_{a,b,c=1}^3\varepsilon_{abc}{\rm e}^{i(k_c-k_a)}C_{abc}(k_1,k_2,k_3),
\end{eqnarray}
and
\begin{equation}
\Phi^{(j)}(s,c,z)=\Phi^{(j)}_0(E,k)+\Phi^{(j)}_1(E,k)s,\qquad j=-,0,+,
\end{equation}
where
\begin{eqnarray}
&&\Phi^{(-)}_0(E,k)=\Big(\frac{J}{2}+(J_z-Jc)c\Big)\sin{k},\nonumber\\
&&\Phi^{(-)}_1(E,k)=\frac{J}{2}+(Jc-J_z)\cos{k},\nonumber\\
&&\Phi^{(0)}_0(E,k)=\frac{3}{2}J_zJc-J_z^2+\frac{J^2}{2}\Big(1-c^2\Big)+\Big[J_zJ\Big(2c^2-\frac{1}{2}\Big)
-J_z^2c\nonumber\\
&&+J^2c(1-c^2)\Big]\cos{k},\nonumber\\
&&\Phi^{(0)}_1(E,k)=\Big[2J_zJc-J_z^2+\frac{J^2}{2}\Big(1-2c^2\Big)\Big]\sin{k},\nonumber\\
&&\Phi^{(+)}_0(E,k)=\frac{J^3}{4}\Big(c+\cos{k}\Big)+\frac{J_z^2J}{2}\Big((2c^2-1)\cos{k}+c\Big)-J_z^3(1+c\cos{k}),
\nonumber\\
&&\Phi^{(+)}_1(E,k)=J_z^2(Jc-J_z)\sin{k}.
\end{eqnarray}
Taking $C_{abc}(k_1,k_2,k_3)$ as a scalar symmetric function
\begin{equation}
C_{abc}(k_1,k_2,k_3)=\Phi^{(+)}(s,c,k),
\end{equation}
we according to Eqs. (23) and (26) readily solve the condition $X^{(+)}(k_1,k_2,k_3)=0$.
Then according to Eqs. (22) and (24) the system (17) reduces to the form
\begin{equation}
\Phi^{(0)}(s,c,k)=0,\qquad
\Phi^{(-)}(s,c,k)=0.
\end{equation}
A substitution of (27) into (30) results in
\begin{eqnarray}
&&\left|\begin{array}{cc}
\Phi^{(0)}_0(E,k)&\Phi^{(0)}_1(E,k)\\
\Phi^{(-)}_0(E,k)&\Phi^{(-)}_1(E,k)
\end{array}\right|=\frac{J^3}{4}\Big(c+\cos{k}\Big)\Big(\cos{k}-4x^3+3x\Big)=0,\quad x=c-\frac{J_z}{J},\qquad\\
&&s=-\frac{\Phi_0^{(-)}(E,k)}{\Phi_1^{(-)}(E,k)}.
\end{eqnarray}
Here (31) is an equation on $E$ and $k$ only while (32) directly defines $s$ for given $E$ and $k$.

From the trigonometric formulas $\cos{3\alpha}=4\cos^3{\alpha}-3\cos{\alpha}$ and
\begin{equation}
c+\cos{k}=4\cos{\frac{k_1+k_2}{2}}\cos{\frac{k_2+k_3}{2}}\cos{\frac{k_3+k_1}{2}},
\end{equation}
follows that Eq. (31) is satisfied only if
\begin{equation}
k_j+k_l=\pm\pi,
\end{equation}
for some $j$ and $l$ or if
\begin{equation}
c-\frac{J_z}{J}=\cos{\Big(\frac{k+2\pi j}{3}\Big)},\qquad j=0,1,2.
\end{equation}

Taking in the case (34) $j=1$, $l=2$ one readily gets
\begin{equation}
c=\cos{k_3}=-\cos{k},\quad\sin{k_2}=\sin{k_1},\quad\sin{k_3}=-\sin{k},\quad s=2\sin{k_1}-\sin{k}.
\end{equation}
Now a substitution of (36) reduces (32) to
\begin{equation}
2\sin{k_1}-\sin{k}=-\sin{k}\Longrightarrow\sin{k_1}=0.
\end{equation}
Hence up to permutations of wave numbers the system (34) has the solution
\begin{equation}
k_1=0,\qquad k_2=\pm\pi.
\end{equation}

In the case (35) we also may exclude all the parameters except for the wave numbers. In fact a substitution of (35) into expressions (28) for $\Phi_j^{(-)}(E,k)$ and then into (32) results in
\begin{equation}
s\Big[1+2\cos{\Big(\frac{k+2\pi j}{3}\Big)}\cos{k}\Big]+\Big[1-2c\cos{\Big(\frac{k+2\pi j}{3}\Big)}\Big]\sin{k}=0,
\qquad j=0,1,2.
\end{equation}
Using the computer algebra system MAPLE this equation may be readily reduced to the form
\begin{equation}
\sin{\frac{k+2\pi(j+1)}{3}}\sin{\frac{k+2\pi(j+2)}{3}}
\Big[\sum_{l=1}^3\sin{\Big(\frac{2(k+2\pi j)}{3}-k_l\Big)}-\sin{\frac{k+2\pi j}{3}}\Big]=0,
\end{equation}
which is solvable either at
\begin{equation}
k=l\pi,\quad l\in{\mathbb Z},
\end{equation}
or at
\begin{equation}
\sum_{l=1}^3\sin{\Big(\frac{2(k+2\pi j)}{3}-k_l\Big)}-\sin{\frac{k+2\pi j}{3}}=0,\qquad j=0,1,2.
\end{equation}
We see that the usual Bethe ansatz is relevant to a very limited class of states.

\subsection{$M=2$ DDD Bethe Ansatz}

If we take the wave function in the $M=2$ DDD Bethe ansatz form \cite{5}
\begin{eqnarray}
&&a_{n_1,n_2,n_3}(k_1,k_2,k_3,\tilde k_1,\tilde k_2,\tilde k_3)=D\varphi(\tilde k_1,\tilde k_2,\tilde k_3)a_{n_1,n_2,n_3}(k_1,k_2,k_3)\nonumber\\
&&+\tilde D\varphi(k_1,k_2,k_3)a_{n_1,n_2,n_3}(\tilde k_1,\tilde k_2,\tilde k_3),
\end{eqnarray}
where $D$ and $\tilde D$ are come coefficients
\begin{equation}
\sum_{j=1}^3E_{magn}(\tilde k_j)=\sum_{j=1}^3E_{magn}(k_j),\quad\sum_{j=1}^3\tilde k_j=\sum_{j=1}^3k_j
\Longleftrightarrow \tilde c=c,\quad\tilde k=k,
\end{equation}
and both the functions $a_{n_1,n_2,n_3}(k_1,k_2,k_3)$ and $a_{n_1,n_2,n_3}(\tilde k_1,\tilde k_2,\tilde k_3)$ are given by (15) and (9), then the system (30) reduces to the form
\begin{equation}
\Phi^{(j)}(s,c,k)D+\Phi^{(j)}(\tilde s,c,k)\tilde D=0,\qquad j=0,-,
\end{equation}
which is solvable only under the condition
\begin{equation}
\left|\begin{array}{cc}
\Phi^{(0)}(s,c,k)&\Phi^{(0)}(\tilde s,c,k)\\
\Phi^{(-)}(s,c,k)&\Phi^{(-)}(\tilde s,c,k)
\end{array}\right|=
\left|\begin{array}{cc}
\Phi^{(0)}_0(E,k)&\Phi^{(0)}_1(E,k)\\
\Phi^{(-)}_0(E,k)&\Phi^{(-)}_1(E,k)
\end{array}\right|(\tilde s-s)=0.
\end{equation}
Since we have assumed that $\tilde s\neq s$ Eq. (46) results in Eq. (31). We see that the set of $M=2$ DDD Bethe states is bigger than the set of usual ($M=1$) Bethe states because now it is not necessary to take into account Eq. (32) and the solvability conditions are limited by Eqs. (34) or (35).

\subsection{$M=3$ DDD Bethe Ansatz}

Taking the wave function in the form
\begin{eqnarray}
&&a_{n_1,n_2,n_3}(k_1^{(1)},k_2^{(1)},k_3^{(1)},k_1^{(2)},k_2^{(2)},k_3^{(2)},k_1^{(3)},k_2^{(3)},k_3^{(3)})
=D_1\varphi(k_1^{(2)},k_2^{(2)},k_3^{(2)})\varphi(k_1^{(3)},k_2^{(3)},k_3^{(3)})\nonumber\\
&&\cdot a_{n_1,n_2,n_3}(k_1^{(1)},k_2^{(1)},k_3^{(1)})+D_2\varphi(k_1^{(3)},k_2^{(3)},k_3^{(3)})
\varphi(k_1^{(1)},k_2^{(1)},k_3^{(1)})a_{n_1,n_2,n_3}(k_1^{(2)},k_2^{(2)},k_3^{(2)})\nonumber\\
&&+D_3\varphi(k_1^{(1)},k_2^{(1)},k_3^{(1)})
\varphi(k_1^{(2)},k_2^{(2)},k_3^{(2)})a_{n_1,n_2,n_3}(k_1^{(3)},k_2^{(3)},k_3^{(3)}),
\end{eqnarray}
where all $a_{n_1,n_2,n_3}(k_1^{(l)},k_2^{(l)},k_3^{(l)})$ are given by Eqs. (15) and (9) and implying conditions analogous to (44) we reduce the system (30) to a solvable system of two equations on three variables
\begin{equation}
\sum_{l=1}^3\Phi^{(j)}(s^{(l)},c,k)D_l=0,\qquad j=-,0.
\end{equation}
When its rank is maximal and equal to two (otherwise we may apply $M=2$ DDD or usual Bethe ansatzes) the solution up to a constant factor has the form
\begin{equation}
D_l=\sum_{a,b=1}^3\varepsilon_{lab}\Phi^{(-)}(s^{(a)},c,k)\Phi^{(0)}(s^{(b)},c,k).
\end{equation}

\section{Summary and discussions}

In the present paper using the standard \cite{1,2,3,4} and the degenerative discrete-diffractive (DDD) \cite{5} versions of Bethe ansatz we studied the three magnon problem for the Hamiltonian (1). We have shown that the very small class of states limited by the conditions (31) and (32) have the usual Bethe form (15). Such "partial integrability" is inherent in some other models \cite{9,10}. The more bigger class of states limited by the single condition (31) may be described with the use of $M=2$ DDD Bethe ansatz (43). Finitely the general case is relevant to the $M=3$ DDD Bethe ansatz (47).

As it was in the case of $S=1$ ferromagnetic chain \cite{5} the space of states obtained with the use of the DDD Bethe ansatz is rude. The latter obstacle however is not fatal if using this system of states we can obtain for the three-magnon sector the "resolution of unity" \cite{11}
\begin{equation}
\sum_{|\mu\rangle}|\mu\rangle\langle\mu|={\cal P}_3,\qquad \hat H|\mu\rangle=E_{\mu}|\mu\rangle,
\end{equation}
where ${\cal P}_3$ is the projection operator on the three-magnon sector. Utilizing this formula one may study a lot of physical applications \cite{12}. Evaluation of such resolution for ${\cal P}_3$ is however an open problem as well as some other ones (four-magnon spectrum and three-magnon scattering) discussed previously \cite{5}.

\end{document}